\documentclass[12pt]{amsart}

\usepackage{natbib} 
\usepackage{amssymb,amsmath,amsfonts}
\usepackage{latexsym,graphicx,color}
\usepackage{mathrsfs} 
\usepackage{dsfont}

\allowdisplaybreaks[4]
\def\cal{\mathcal} 

\def\shuffle{{\, \shuffl \,}}

\def\unitstep{\mathbb U}

\definecolor{Light}{gray}{0.85}

\def\abs#1{\left\vert #1 \right\vert}
\def\allpoly{\mbox{$\re\langle X \rangle$}}
\def\allpolyell{\mbox{$\re^{\ell}\langle X \rangle$}}
\def\allpolyx0degn{\mbox{$P_n$}}
\def\allpolyXO{\mbox{$\re [X_0]$}}

\def\allseries{\mbox{$\re\langle\langle X \rangle\rangle$}}

\def\allseriesell{\mbox{$\re^{\ell} \langle\langle X \rangle\rangle$}}
\def\allseriesGC{\mbox{$\re_{GC}\langle\langle X \rangle\rangle$}}
\def\allseriesLC{\mbox{$\re_{LC}\langle\langle X \rangle\rangle$}}

\def\allseriesellLC{\mbox{$\re^{\ell}_{LC}\langle\langle X \rangle\rangle$}}
\def\allseriesellGC{\mbox{$\re^{\ell}_{GC}\langle\langle X \rangle\rangle$}}

\def\bfem#1{{\bf \em #1}} 

\def\bull{\rule{0.08in}{0.08in}} 

\def\C{{\mathbb C}} 

\def\charseries{{\rm char}}

\newcommand{\comment}[1]{} 

\def\dim{{\rm dim}}

\def\eqref#1{(\ref{#1})} 

\def\Lpm{L_{\mathfrak{p}}^m}

\def\Lpme{L^m_{\mathfrak{p},e}}

\def\liead{{\rm ad}}

\def\liepoly{{\cal L}(X)}

\def\nat{{\mathbb N}} 
\def\norm#1{\left\Vert#1\right\Vert}

\def\openbull{\framebox[0.08in][c]{$\;$}} 

\def\re{{\mathbb R}} 

\def\shuffle{{\scriptscriptstyle \;\sqcup \hspace*{-0.05cm}\sqcup\;}}

\def\supp{{\rm supp}}

\newtheorem{mylemma}{Lemma}[section]
\newtheorem{mydefinition}{Definition}[section]
\newtheorem{mytheorem}{Theorem}[section]
\newtheorem{mycorollary}{Corollary}[section]
\newtheorem{myexample}{Example}[section]

\def\begals{\[\begin{aligned}}
\def\endals{\end{aligned}\]}
\def\begce{\begin{center}}
\def\endce{\end{center}}
\def\begar{\begin{array}}
\def\endar{\end{array}}
\def\begeq{\begin{equation}}
\def\endeq{\end{equation}}
\def\begdi{\begin{displaymath}}
\def\enddi{\end{displaymath}}
\def\begdis{\begin{eqnarray*}}
\def\enddis{\end{eqnarray*}}
\def\begeqa{\begin{eqnarray}}
\def\endeqa{\end{eqnarray}}
\def\begdes{\begin{description}}
\def\enddes{\end{description}}
\def\begit{\begin{itemize}}
\def\endit{\end{itemize}}
\def\begen{\begin{enumerate}}
\def\enden{\end{enumerate}}
\def\beglar{\left[\begin{array}}
\def\endrar{\end{array}\right]}
\def\begle{\begin{mylemma}}
\def\endle{\end{mylemma}}
\def\begde{\begin{mydefinition}}
\def\endde{\end{mydefinition}}
\def\begth{\begin{mytheorem}}
\def\endth{\end{mytheorem}}
\def\begco{\begin{mycorollary}}
\def\endco{\end{mycorollary}}
\def\begprop{\begin{myproposition}}
\def\endprop{\end{myproposition}}
\def\begex{\begin{myexample} \rm}
\def\endex{\hfill\openbull \end{myexample} \vspace*{0.15in}}
\def\begexer{\begin{myexercise}}
\def\endexer{\end{myexercise}}

\def\begres{\noindent{\bf Remarks}:\begin{enumerate}}
\def\endres{\end{enumerate} \par}
\def\begpr{\noindent{\em Proof:}$\;\;$}
\def\endpr{\hfill\bull \vspace*{0.15in}}
\def\begtab{\begin{tabular}}
\def\endtab{\end{tabular}}
\def\rref#1{(\ref{#1})}

\begin{document}


\title[On Symmetries in Analytic Input-Output Systems]
{On Symmetries in Analytic Input-Output Systems}

\author{W. Steven Gray}
\address{Department of Electrical and Computer Engineering, Old Dominion University, Norfolk, Virginia 23529, USA}
\email{sgray@odu.edu}
\urladdr{http://www.ece.odu.edu/~sgray}
\date{}

\author{Erik I. Verriest}
\address{Department of Electrical and Computer Engineering, Georgia Institute of Technology, Atlanta, GA  30332}
\email{erik.verriest@ece.gatech.edu}
\date{}

\begin{abstract}
There are many notions of symmetry for state space models. They play a role in understanding when systems are time reversible,
provide a system theoretic interpretation of thermodynamics, and have applications in certain stabilization and optimal control problems.
The earliest form of symmetry for analytic input-output systems is due to Fliess who introduced systems described by an {\em exchangeable} generating series. In this case, one is able to write the output as a memoryless
analytic function of the integral of each input.
The first goal of this paper is to describe two new types of symmetry for such Chen--Fliess input-output systems, namely,
{\em coefficient reversible symmetry} and {\em palindromic symmetry}. Each concept is then related to the notion of an exchangeable series.
The second goal of the paper is to provide an in-depth analysis of Chen--Fliess input-output systems whose
generating series are linear time-varying, palindromic, and have generating series coefficients growing at a maximal rate while ensuring some type of convergence. It is shown that such series have an infinite
Hankel rank and Lie rank, have a certain infinite dimensional state space realization, and a description of their relative degree and zero dynamics is given.
\end{abstract}

\maketitle

\thispagestyle{empty}

\noindent
\tableofcontents


\vspace{0.1in}


\section{Introduction}

Symmetry in the context of control theory has a long history beginning with the work of Brockett and (J.~L.)~Willems
who demonstrated that linear time-invariant (LTI) systems with block circulant symmetry appear naturally in lumped approximations of
distributed systems \citep{Brockett-Willems_74}. Subsequently,
many different notions of symmetry have emerged depending on the underlying model involved and the application being pursed.
Roughly speaking, the existing definitions break down into concepts best suited for LTI systems and those aimed at more general classes of systems

In the context of LTI systems, there was the early work of (J.~C.)~Willems who introduced the behavioral approach to
system modeling and defined the property of {\em time-reversal symmetry} as one where the time-reversed version of any input-output pair of
a given system is also an admissible input-output pair for the system \citep{Willems_78}. The question at the time was what type of structure does this
kind of symmetry impose on any finite dimensional linear state space realization of the input-output map.
These ideas were further developed by \cite{Fagnani-Willems_91} and later expanded to other forms of symmetry using transformations groups
\citep{Fagnani-Willems_93,Fagnani-Willems_94}. A second notion of time-reversal symmetry was initiated by \cite{Bernstein-Bhat_03}
who considered {\em output reversal symmetry} of the free response of a linear dynamical system with an output map. This property is characterized
by the feature that if an output is in the range of the initial state-to-output map, then its time-reversed version is also in the range of this map.
The aim here was to provide a system theoretic interpretation of thermodynamics. This program was later fully realized by
\cite{Haddad-etal_05,Haddad-etal_08,Nersesov-Haddad_08,Nersesov-etal_12,Nersesov-etal_14}.
A third type of symmetry appearing in the LTI case is that of palindromic symmetry in system parameterizations.
This is defined in terms of palindromic polynomials appearing in the transfer function \citep{Butkovskii_94}. This leads to yet another definition of
time-reversibility \citep{Markovsky-Rao_08} and turns out to be useful, for example, in certain types of stabilization problems \citep{Volinsky-Shklyar_23}.
A final kind of symmetry in the LTI setting is that of algebraic symmetry
\citep{Martin_82, Hazewinkel-Martin_83,Hazewinkel-Martin_84}. The main idea here is that systems composed of identical interacting subsystems have an
underlying {\em symmetry algebra}. This structure allows one to define a system over an algebra in a natural way so that problems like stabilization can
be solved while preserving the intrinsic structure of the plant.

The earliest notion of symmetry for more general classes of systems is due to Fliess
in the context of analytic input-output systems written in terms of weighted sums
of iterated integrals, what is now called a {\em Chen--Fliess series} \citep{Fliess_81,Fliess_83}.
The weights or coefficients of such a functional series are indexed by words over a finite alphabet and
define a noncommutative formal power series or generating series. A generating series is said to be {\em exchangeable}
if every coefficient is invariant under any permutation of the letters in its index (see Definition~\ref{de:exchangeable-series}) \citep{Fliess_74,Fliess_81}.
This type of symmetry for generating series is equivalent to
being able to write its corresponding Chen--Fliess series as a memoryless analytic function of the integral of each input.
Other lines of research in this area provide notions of symmetry in the state space setting.
For physical systems constrained by conservation laws, it was shown by \cite{van-der-Schaft_81,van-der-Schaft_83,van-der-Schaft_84,van-der-Schaft_87} that notions of symmetry
can be used to simplify and solve optimal control problems.
Conversely, conditions can be identified for
Hamiltonian systems under which they possess time-reversal symmetry in the sense of \cite{Willems_78}.
More general types of state space symmetry were developed by
\cite{Grizzle-Marcus_84,Grizzle-Marcus_85} in terms of symmetry groups acting on realizations in order to
decompose the system into lower dimensional subsystems. The motivation was to simplify the
analysis of the original system by investigating the properties of these simpler subsystems.

The first goal of this paper is to describe two new types of symmetry for Chen--Fliess input-output systems, namely,
{\em coefficient reversible symmetry} and {\em palindromic symmetry}. Coefficient reversible symmetry refers to systems whose
generating series are invariant when the indices of the coefficients are written in reverse order. This concept is
close in spirit to palindromic polynomials used in the analysis of LTI systems but involves a noncommuting
alphabet. It is shown that such systems have a type
of {\em input reversal symmetry}. That is, the output value at time $t$ is unchanged when the input is time-reversed on
the interval $[0,t]$.
Palindromic symmetry refers to the subclass of coefficient reversible series with the defining property
that every word in the support of a given generating series is a palindrome.
LTI systems never possess this type of symmetry (see Example~\ref{ex:LTI-palindromic-counter-example}).
Palindromic words have a long history in theoretical computer science \citep{Droubay-Pirillo_99}. More recently, DNA palindromes in the human genome have been identified as playing a role in carcinogenesis \citep{Miklenic-Svetec_21}. Their presence in the SARS-CoV-2 virus was used as a rapid detection tool to
track the evolution of the virus during the pandemic \citep{Ghosh-etal_22}.
Palindromes are also
used in the genome editing technology CRISPR (Clustered Regularly Interspaced Short Palindromic Repeats) \citep{Jinek-etal_12}.
But their appearance in the context of system theory appears to be new.
Finally, these two types of series symmetry are related to the existing and distinct notion of exchangeable series.

The second goal of the paper is to provide an in-depth analysis of Chen--Fliess input-output systems whose
generating series are linear time-varying and palindromic.
A specific instance of this type of generating series
was used by \cite{Gray-Verriest_23} to show that a class of functional differential equations are well suited for describing such systems.
That example is generalized here in
Sections \ref{sec:GC-nonexistent-finite-realizations} and \ref{sec:Infinite-Dimensional-State-Space-Realization} to series that
are {\em globally maximal} in a sense to be described.
It is shown that their generating series are algebraic but are neither rational nor possess finite Lie rank.
Nevertheless, the input-output map does have an infinite dimensional, linear time-varying state space realization in terms of
Bessel functions of the first kind.
The corresponding impulse response can be described in terms of a modified Bessel function of the first kind.
The problem of zeroing the output of this operator is then addressed when an arbitrary zero-input response is added to the output.
By purely algebraic means it is shown that the {\em nulling} input can be written in terms of a Neumann series of Bessel functions \citep{Jankov-etal_11,Wilkins_48}.
Once this input is identified, it is shown that the state space realization has relative degree one, and the zero dynamics are explicitly described \citep{Isidori_95}.
Next, a class of palindromic linear systems that are convergent only in a local sense are characterized. The same three problems are addressed, namely, the
Hankel rank and Lie rank of the generating series are shown to be infinite, an infinite dimensional state space realization is presented, and a description of the relative degree and zero dynamics of such systems are given. These problems are more difficult since the impulse response exhibits singularities, and the role of the Bessel functions is superseded by generalized hypergeometric functions.

The paper is organized as follows. The next section summarizes some mathematical preliminaries to establish the notation and terminology.
Section~\ref{sec:classes-of-symmetric-FPS} then presents the two new notions of symmetry for (generating) formal power series.
The characterization of globally maximal palindromic linear systems is developed in Section~\ref{sec:Globally Maximal Palindromic Linear Systems}.
Section~\ref{sec:Locally Maximal Palindromic Linear Systems} presents the analogous results for the local case. The conclusions of the paper are
given in the final section.

\section{Preliminaries}
\label{sec:preliminaries}

An {\em alphabet} $X=\{ x_0,x_1,$ $\ldots,x_m\}$ is any nonempty finite set
of symbols referred to as {\em
letters}. A {\em word} $\eta=x_{i_1}x_{i_2}\cdots x_{i_k}$ is a finite sequence of letters from $X$.
The number of letters in a word $\eta$, written as $\abs{\eta}$, is called its {\em length}.
The empty word, $\emptyset$, has length zero.
The collection of all words is denoted by $X^\ast$ and constitutes
a noncommutative monoid under concatenation.
Given any nonempty word $\eta$, let $\eta X^\ast$ be the set of all words with the prefix $\eta$.
Any mapping $c:X^\ast\rightarrow
\re^\ell$ is called a {\em formal power series}.
It is often
written as the formal sum $c=\sum_{\eta\in X^\ast}( c,\eta)\eta$,
where the {\em coefficient} $(c,\eta)\in\re^\ell$ is the image of
$\eta\in X^\ast$ under $c$.
The {\em support} of $c$, $\supp(c)$, is the set of all words having nonzero coefficients.
A series is {\em proper} if $\supp(c)$ does not contain the empty word.
The set of all noncommutative formal power series over the alphabet $X$ is
denoted by $\allseriesell$. The subset of series with finite support, i.e., polynomials,
is represented by $\allpolyell$.

\subsection{Chen--Fliess series}
\label{subsec:CF-series}

Given any $c\in\allseriesell$, one can associate a
functional series, $F_c$, in the following manner.
Let $\mathfrak{p}\ge 1$ and $t_0 < t_1$ be given. For a Lebesgue measurable
function $u: [t_0,t_1] \rightarrow\re^m$, define
$\norm{u}_{\mathfrak{p}}=\max\{\norm{u_i}_{\mathfrak{p}}: \ 1\le
i\le m\}$, where $\norm{u_i}_{\mathfrak{p}}$ is the usual
$L_{\mathfrak{p}}$-norm for a measurable real-valued function,
$u_i$, defined on $[t_0,t_1]$.  Let $L^m_{\mathfrak{p}}[t_0,t_1]$
denote the set of all measurable functions defined on $[t_0,t_1]$
having a finite $\norm{\cdot}_{\mathfrak{p}}$ norm and
$B_{\mathfrak{p}}^m(R)[t_0,t_1]:=\{u\in
L_{\mathfrak{p}}^m[t_0,t_1]:\norm{u}_{\mathfrak{p}}\leq R\}$.
Assume $C[t_0,t_1]$
is the subset of continuous functions in $L_{1}^m[t_0,t_1]$. Define
inductively for each word $\eta=x_i\bar{\eta}\in X^{\ast}$ the map $E_\eta:
L_1^m[t_0, t_1]\rightarrow C[t_0, t_1]$ by setting
$E_\emptyset[u]=1$ and letting
\[E_{x_i\bar{\eta}}[u](t,t_0) =
\int_{t_0}^tu_{i}(\tau)E_{\bar{\eta}}[u](\tau,t_0)\,d\tau, \] where
$x_i\in X$, $\bar{\eta}\in X^{\ast}$, and $u_0=1$. The
{\em Chen--Fliess series} corresponding to $c\in\allseriesell$ is
\begeq
y(t)=F_c[u](t) =
\sum_{\eta\in X^{\ast}} ( c,\eta) \,E_\eta[u](t,t_0) \label{eq:Fliess-operator-defined}
\endeq
\citep{Fliess_81}.
To establish the convergence of this series, assume
there exist real numbers $K,M>0$ such that
\begin{equation}
\abs{(c,\eta)}\le K M^{|\eta|}|\eta|!,\;\; \forall\eta\in X^{\ast}.
\label{eq:local-convergence-growth-bound}
\end{equation}
(Define $\abs{z}=\max_i \abs{z_i}$ whenever $z\in\re^\ell$.)
It is shown by \cite{Gray-Wang_02}
that under such circumstances
the series \rref{eq:Fliess-operator-defined} converges uniformly and
absolutely so that $F_c$ describes a well defined mapping from
$B_{\mathfrak p}^m(R)[t_0,$ $t_0+T]$ into $B_{\mathfrak
q}^{\ell}(S)[t_0, \, t_0+T]$ for sufficiently small $R,T>0$, where
$\mathfrak{p},\mathfrak{q}\in[1,\infty]$ satisfy $1/\mathfrak{p}+1/\mathfrak{q}=1$.
The operator $F_c$ is said to be {\em locally convergent} and
is called a {\em Fliess operator}.
The collection of all generating
series $c$ satisfying
the growth condition \rref{eq:local-convergence-growth-bound}
is denoted by $\allseriesellLC$.
A series $c\in\allseriesLC$ is called {\em locally maximal} when it has the form
$c=\sum_{\eta\in X^\ast} KM^{|\eta|}|\eta|!\,\eta$ (see Definition~\ref{de:maximal-series}).
It provides a lower bound on
the radius of convergence as described by \cite{Thitsa-Gray_SIAM12} for any Fliess operator
whose coefficients satisfy \rref{eq:local-convergence-growth-bound}.
When $c$ complies with the more restrictive growth condition
\begeq
\abs{(c,\eta)}\le K M^{|\eta|}(|\eta|!)^s,\;\; \forall\eta\in X^{\ast}, \label{eq:global-convergence-growth-bound}
\endeq
with $s\in[0,1)$,
the series \rref{eq:Fliess-operator-defined}
defines an input-output operator from the $L_p$ extended space
\begdi
\Lpme(t_0):=
\{u:[t_0,\infty)\rightarrow \re^m:u_{[t_0,t_1]}\in \Lpm[t_0,t_1],\;\;\forall t_1 \in (t_0,\infty)\}
\enddi
into $C[t_0, \infty)$,  where
$u_{[t_0,t_1]}$ is taken as the restriction of the input $u$ to the interval $[t_0,t_1]$ \citep{Winter-Arboleda-etal_15}.
The operator $F_c$ in this case is called {\em globally convergent}, and the set of all generating
series with a growth bound \rref{eq:global-convergence-growth-bound}
is denoted by $\allseriesellGC$.
A series $c\in\allseriesGC$ will be called {\em globally maximal} when
$c=\sum_{\eta\in X^\ast} KM^{|\eta|}\,\eta$.

\subsection{Differential representations}

A series $c\in\allseriesell$ is said to have a {\em differential representation} when
there exists a $z_0\in\re^n$, a smooth function $h:\re^n\rightarrow \re^\ell$ defined on a neighborhood $W$ of
$z_0$, and an $m$-tuple of smooth vector fields $(g_0,g_1,\ldots,g_m)$ defined on $W$ such that
for any word $\eta=x_{i_k}\cdots x_{i_1}\in X^\ast$
\begdi
(c_j,\eta)=L_{g_{\eta}}h_j(z_0):=L_{g_{i_1}}\cdots L_{g_{i_k}}h_j(z_0),
\enddi
$j=1,2,\ldots,\ell$,
where $L_{g_i}h_j$ is the Lie derivative of $h_j$ with respect to $g_i$. Such a
representation can also be only formal when all the functions involved are formal, i.e., defined by
possibly nonconvergent Taylor series at $z_0$. In which case, the derivatives are interpreted as left-shifts on
the Taylor series, and the Chen--Fliess series is a formal sum.
In either case, the input-output map $y=F_c[u]$ has a control-affine state space realization
\begin{align*}
\dot{z}&=g_0(z)+\sum_{i=1}^m g_i(z)u_i,\;\;z(0)=z_0 \\
y&=h(z).
\end{align*}
The existence of a differential representation is
established using the following concept.

\begde \citep{Fliess_74}
Given any $c\in\allseriesell$, the $\re$-linear mapping
${\mathcal H}_c:\allpoly\rightarrow \allseriesell$
uniquely specified by
$
({\mathcal H}_c(\eta),\xi)=(c,\xi\eta)$, $\forall \xi,\eta \in X^{\ast}
$
is called the \bfem{Hankel mapping} of $c$.
\endde

A series $c\in\allseriesell$ is said to
have finite {\em Lie rank} $\rho_L(c)$ when the vector space ${\mathcal H}_c(\liepoly)$ has dimension
$\rho_L(c)$, where $\liepoly\subset \allpoly$ is the free Lie algebra over $X$.
It is well known that $c$ has a differential representation if and
only if $c$ has finite Lie rank
\citep{Fliess_81,Fliess_83,Isidori_95,Jakubczyk_80,Jakubczyk_86a,Jakubczyk_86b,Sussmann_77,Sussmann_90}.
All minimal representations
have dimension $\rho_L(c)$ and are unique up to a diffeomorphism.

In the event that $c\in\allseries$ and the vector space ${\mathcal H}_c(\allpoly)$ has finite dimension $\rho_H(c)$,
then $c$ has {\em Hankel rank} $\rho_H(c)$ and a
differential representation where all the vector fields are linear. That is,
\begeq \label{eq:linear-representation}
(c,x_{i_k}\cdots x_{i_1})=CN_{i_k}\cdots N_{i_1}z_0, 
\endeq
where $C$, $N_i$, and $z_0$ are matrices with real coefficient and of dimension $1\times\rho_H(c)$, $\rho_H(c)\times\rho_H(c)$, and $\rho_H(c)\times 1$, respectively \citep{Fliess_74,Fliess_81,Isidori_95}.
It is well known that a series $c\in\allseries$ is {\em rational} if and only if it has a linear representation \rref{eq:linear-representation}
\citep{Berstel-Reutenauer_88,Fliess_81,Salomaa-Soittola_78}.
In addition, any system $y=F_c[u]$ with $c$ rational has a bilinear realization
\begin{align*}
\dot{z}&=N_0(z)+\sum_{i=1}^m N_i(z)u_i,\;\;z(0)=z_0 \\
y&=Cz.
\end{align*}

\subsection{Algebraic formal power series}
\label{subsec:algebraic-FPS}

A generalization of rational series is the class of algebraic series.
Such series are characterized by being a solution to a system of
polynomial equations as described below.

\begde \label{de:algebraic-series} \citep{Salomaa-Soittola_78,Schutzenberger_62}
Let $Z=\{z_1,z_2,\ldots,z_n\}$ be an alphabet disjoint from alphabet $X$. A \bfem{proper $\re$-algebraic system} is a set of equations
$z_i=p_i$, $i=1,2,\ldots,n$ such that:
\begen
\item $p_i\in\re\langle X\bigcup Z \rangle$
\item  $(p_i,\emptyset)=0$ and $(p_i,z_j)=0$, $\forall i,j$.
\enden
\endde

A strong solution to $(p_1,p_2,\ldots,p_n)$ is an $n$-tuple of proper series $c=(c_1,c_2,\ldots,c_n)\in\allseries^n$
computed inductively from the given set of equations
such that
$
c_i=p_i(X,Z)|_{Z_i=c_i}.
$
Each $c_i$ is called a {\em component} of $c$.
Every proper $\re$-algebraic system is known to have a unique solution.
A series $d\in\allseries$ is called {\em $\re$-algebraic} if its proper part $d-(d,\emptyset)$ is a component of a
proper $\re$-algebraic system.
Every rational series is a solution to a proper $\re$-algebraic system of {\em linear} equations.
Thus, algebraic series and differentially generated series share a common subset, i.e., rational series.
A representation theory for algebraic series is well known \citep{Shamir_67,Petre-Salomaa_09}, but it will not
be so useful in the present context.

\section{Three classes of symmetric formal power series}
\label{sec:classes-of-symmetric-FPS}

Three classes of formal power series are defined in this section, each with a certain kind of symmetry:
coefficient reversible series, palindromic series, and exchangeable series.

Let $X=\{ x_0,x_1,$ $\ldots,x_m\}$ and
define the following involution on the set of words $X^\ast$:
\begdi
\sim:\eta=x_{i_1}x_{i_2}\cdots x_{i_k}\in X^\ast\mapsto \tilde{\eta}=x_{i_k}x_{i_{k-1}}\cdots x_{i_1}.
\enddi
Linearly extend this map to $\allseries$ so that $\tilde{c}=\sum_{\eta\in X^\ast} (c,\eta)\tilde{\eta}$.

\begde
A series $c\in\allseries$ is \bfem{coefficient reversible} if
\begdi
(c,\eta)=(c,\tilde{\eta}),\;\;\forall \eta\in X^\ast.
\enddi
\endde

\begex
The polynomial $c=x_0x_1+x_1x_0$ is coefficient reversible while $d=x_0x_1$ is not.
\endex

\begle \label{le:coefficient-reversible}
A series is coefficient reversible if and only if $\tilde{c}=c$.
\endle

\begpr
Suppose $c=\sum_{\eta\in X^\ast}(c,\eta)\eta$ is coefficient reversible. Then
\begdi
\tilde{c}=\sum_{\eta\in X^\ast}(c,\eta)\tilde{\eta}=\sum_{\eta\in X^\ast}(c,\tilde{\eta})\tilde{\eta}
=\sum_{\tilde{\eta}\in X^\ast}(c,\tilde{\eta})\tilde{\eta}=c
\enddi
using the fact that $\sim$ is bijective.
Conversely, if $\tilde{c}=c$, then directly
\begdi
\sum_{\eta\in X^\ast}(c,\eta)\tilde{\eta}=\sum_{\eta\in X^\ast}(c,\eta)\eta=\sum_{\tilde{\eta}\in X^\ast}(c,\tilde{\eta})\tilde{\eta}
=\sum_{\eta\in X^\ast}(c,\tilde{\eta})\tilde{\eta},
\enddi
which implies that
\begdi
(c,\eta)=(c,\tilde{\eta}),\;\;\forall \eta\in X^\ast.
\enddi
\endpr

\begde
An input-output map $F:u\mapsto y$ defined on $[0,T]$ is called \bfem{input reversible} if for every $t\in[0,T]$
\begdi
y(t)=F[u](t)=F[u_{t}](t),
\enddi
where $u_{t}(\tau)=u(t-\tau)$ on $[0,t]$.
\endde

\begle
For $c\in\allseriesLC$ and admissible $u$, $F_c[u](t)=F_{\tilde{c}}[u_t](t)$ for every $t$ in the interval
of convergence [0,T].
\endle

\begpr
For any $t\in[0,T]$ and admissible $u$ observe
\begdi
F_{\tilde{c}}[u_t](t)=\sum_{\eta\in X^\ast} (c,\eta)E_{\tilde{\eta}}[u_t](t,0).
\enddi
The assertion is that $E_{\tilde{\eta}}[u_t](t,0)=E_\eta[u](t,0)$, which would
prove the lemma.
The claim is trivial when $\eta=\emptyset$.
In the case where $\eta=x_{i_1}x_{i_{2}}\cdots x_{i_k}$, observe
\begin{align*}
E_{\eta}[u](t,0)
&=E_{x_{i_1}x_{i_{2}}\cdots x_{i_k}}[u](t,0) \\
&=\int_0^{t}\int_0^{\tau_1}\cdots \int_0^{\tau_{k-1}} u_{i_1}(\tau_1)u_{i_2}(\tau_2)\cdots
u_{i_k}(\tau_{k})\,d\tau_k d\tau_{k-1}\cdots d\tau_1 \\
&=\int_0^{t}\int_0^{t}\cdots \int_0^t u_{i_1}(\tau_1)u_{i_2}(\tau_2)\cdots u_{i_k}(\tau_k)
\unitstep(\tau_1-\tau_2) \unitstep(\tau_2-\tau_3)\cdots \unitstep(\tau_{k-1}-\tau_k) \\
&\hspace*{0.2in} d\tau_k d\tau_{k-1}\cdots d\tau_1,
\end{align*}
where $\unitstep$ denotes the unit step function.
Interchanging the order of integration gives
\begdi
E_{\eta}[u](t,0)=\int_0^{t}\int_{\tau_k}^{t}\cdots \int_{\tau_{2}}^{t}
u_{i_1}(\tau_1)u_{i_{2}}(\tau_{2})\cdots
u_{i_k}(\tau_k)\,d\tau_1d\tau_2\cdots d\tau_{k}.
\enddi
Finally, substituting $t-\tau_1$ for $\tau_1$ followed by
$t-\tau_2$ for $\tau_2$, etc., yields the
desired result, namely,
\begin{align*}
E_{\eta}[u](t,0)
&=\int_0^{t}\int_0^{\tau_k}\cdots\int_0^{\tau_2}
(u_{i_1}(t-\tau_1))(u_{i_2}(t-\tau_2))\cdots
(u_{i_k}(t-\tau_k))\,d\tau_1d\tau_2\cdots d\tau_k \\
&=E_{\tilde{\eta}}[u_{t}](t,0).
\end{align*}
\endpr

The following theorem is immediate.

\begth
If $c\in\allseriesLC$ is coefficient reversible, then $F_c$ is input reversible.
\endth

\begde
A word $\eta\in X^\ast$ is a palindrome if $\tilde{\eta}=\eta$.
A series $c\in\allseries$ is \bfem{palindromic} if every word in its support is a palindrome.
\endde

In addition, $c$ is {\em even} palindromic when it is palindromic and every word in its support has even length.
Likewise, $c$ is {\em odd} palindromic if it is palindromic and every word in its support has odd length.

\begex
The polynomial $c=x_1x_0+x_0x_1$ is coefficient reversible but not palindromic, while
$d=x_0x_1x_0$ is coefficient reversible and odd palindromic.
\endex

\begle
A palindromic series $c$ is coefficient reversible.
\endle

\begpr
If $c$ is palindromic, then
\begdi
\tilde{c}=\sum_{\eta\in X^\ast}(c,\eta)\tilde{\eta}=\sum_{\eta\in\supp(c)}(c,\eta)\eta=\sum_{\eta\in X^\ast}(c,\eta)\eta=c.
\enddi
Hence, from Lemma~\ref{le:coefficient-reversible}, $c$ is coefficient reversible.
\endpr

For any $x_i\in X$ and $\eta\in X^\ast$, let $\abs{\eta}_{x_i}$ denote the number of times
the letter $x_i$ appears in the word $\eta$.

\begde \label{de:exchangeable-series} \cite{Fliess_74,Fliess_81}
A series $c\in\allseries$ is said to be \bfem{exchangeable} if
for all $i\in\{0,1,\ldots,m\}$ and $\eta,\xi\in X^\ast$:
\begdi
\abs{\eta}_{x_i}=\abs{\xi}_{x_i}\Rightarrow (c,\eta)=(c,\xi).
\enddi
\endde

\begex
The polynomial $c=x_1x_0+x_0x_1$ is exchangeable and coefficient reversible, while
$d=x_0x_1x_0$ is  coefficient reversible but not exchangeable.
\endex

\begle
An exchangeable series $c$ is coefficient reversible.
\endle

\begpr
Given any $\eta\in X^\ast$ it follows that $\abs{\eta}_{x_i}=\abs{\tilde{\eta}}_{x_i}$
for all $i\in\{0,1,\ldots,m\}$. If $c$ is exchangeable, then
$(c,\eta)=(c,\tilde{\eta})$. Hence, $c$ is coefficient reversible.
\endpr

\begex
The series $c=x_1x_0+x_0x_1$ is exchangeable but not palindromic.
The series $d=x_0x_1x_0$ is palindromic but not exchangeable.
The series
\begeq \label{eq:P-and-E-series}
e=(e,\emptyset)\emptyset+\sum_{j=1}^\infty (e,x_{i_j}^j) x_{i_j}^j,
\endeq
where $i_j\in\{0,1,\ldots,m\}$,
is both palindromic and exchangeable.
\endex

A series $c\in\allseriesLC$ is exchangeable if only if $F_c$ is an analytic function of $E_{x_i}$, $i=0,1,\ldots,m$ \cite[Proposition II.9]{Fliess_81}.
This implies that $y=F_c[u]$ has a state space realization of the form $\dot{z}_i=u_i$, $z_i(0)=0$, $i=0,1,\ldots,m$,
and $y=h(z)$ with $h$ being real analytic about the origin.
Therefore, $c$ has Lie rank $\rho_L(c)\leq m+1$. The following class of exchangeable generating series
were used by \cite{Thitsa-Gray_SIAM12} to define the notion of radius of convergence for the class of
series whose minimum growth rate is parameterized by $K,M$ in the sense of \rref{eq:local-convergence-growth-bound}
or \rref{eq:global-convergence-growth-bound} (with $s=0$).

\begde \label{de:maximal-series}
A series $c\in\allseriesLC$ is said to be \bfem{locally maximal}
with growth constants $K,M>0$ if $(c,\eta)=K M^{|\eta|}|\eta|!$, $\forall\eta \in X^\ast$.
Likewise, a series $c\in\allseriesGC$ is said to be \bfem{globally maximal}
with growth constants $K,M>0$ if $(c,\eta)=KM^{|\eta|}$, $\forall\eta \in X^\ast$.
\endde

\begex
The assertion is that maximal series always have Lie rank equal to one. But first observe that
the realization of Fliess provides a realization of $y=F_c[u]$ when $c$ is a locally maximal series. Assume $X=\{x_0,x_1,\ldots,x_m\}$ and set
$\dot{z}_i=u_i$, $z_i(0)=0$, $i=0,1,\ldots,m$ with $h(z_1,z_2,\ldots,z_m)=K/(1-Mz_1-Mz_2-\cdots -Mz_m)$.
Then if follows that
\begin{align*}
y&=\frac{K}{1-M\sum_{i=0}^m E_{x_i}[u]} \\
&=K\sum_{k=0}^\infty M^k F^k_{\charseries(X)}[u] \\
&=K\sum_{k=0}^\infty M^k F_{(\charseries(X))^{\shuffle k}}[u],
\end{align*}
where $\charseries(X):=\sum_{i=0}^m x_i$ is the characteristic series of $X$, and
$d^{\shuffle k}$ denotes the shuffle power of $d\in\allseries$ \citep{Fliess_81}. Next apply the identity
$(\charseries(X))^{\shuffle k}=k!(\charseries(X))^k$ and the fact that $(\charseries(X))^k=\sum_{\eta\in X^k}\eta$,
where $X^k$ is the set of all words in $X^\ast$ of length $k$. In which case,
\begin{align*}
y&=K\sum_{k=0}^\infty M^kk!\, F_{(\charseries(X))^k}[u] \\
&=\sum_{k=0}^\infty \sum_{\eta\in X^k} KM^k k!\, E_{\eta}[u] \\
&=\sum_{\eta\in X^\ast} KM^{\abs{\eta}}\abs{\eta}!\,E_{\eta}[u].
\end{align*}
Therefore, $c=\sum_{\eta\in X} KM^{\abs{\eta}}\abs{\eta}!\,\eta$ as claimed.
In particular, note that this realization is not minimal since the realization
\begdi
\dot{z}=1+\sum_{i=1}^m u_i,\;\;z(0)=0,\;\;y=\frac{K}{1-Mz}
\enddi
yields the same locally maximal generating series. That is, $\rho_L(c)=1$.
A similar calculation holds for the globally maximal case, but instead
yields the minimal realization
\begdi
\dot{z}=1+\sum_{i=1}^m u_i,\;\;z(0)=0,\;\;y=K{\rm e}^{Mz}.
\enddi
\endex

\begin{figure}[tb]
\begin{center}
\includegraphics[scale=0.35]{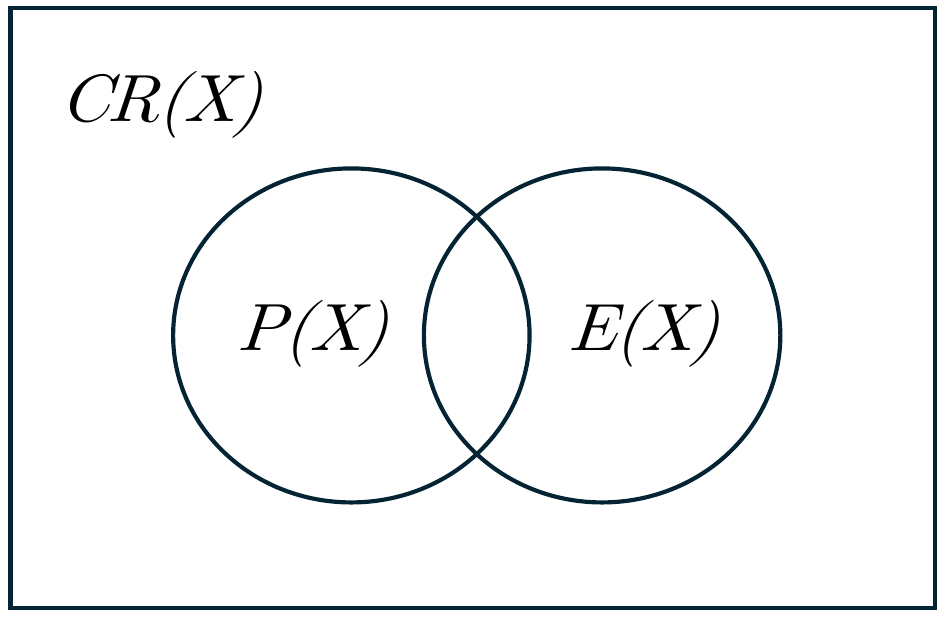}
\caption{Relationship between the classes of symmetric formal power series $CR(X)$, $P(X)$ and $E(X)$.}
\label{fig:symmetric-FPS}
\end{center}
\end{figure}

Let $CR(X)$, $P(X)$ and $E(X)$ be the subsets of $\allseries$ consisting of all coefficient reversible series, palindromic series, and exchangeable series, respectively.
The relationship between these three notions of symmetry for a formal power series is summarized in Figure~\ref{fig:symmetric-FPS}.
The intersection between palindromic and exchangeable series is described in the following theorem.

\begth
A series $e\in P(X)\cap E(X)$ if and only if $e$ has the form \rref{eq:P-and-E-series}.
\endth

\begpr
The only nontrivial claim is that if series $e\in P(X)\cap E(X)$, then it must have the form \rref{eq:P-and-E-series}.
The proof is by contradiction.
Let $e\in P(X)\cap E(X)$ with at least one word $\eta=x_{i_1}x_{i_2}\cdots x_{i_j}$ in its support of length $j\geq 2$ (otherwise the claim is immediate).
If $\eta\neq x_{i_j}^j$ for some $x_{i_j}\in X$, then it contains at least two distinct letters $x_{i_j}\neq x_{i_k}$. Since the series is exchangeable $(e,x_{ij}\eta^\prime x_{i_k})\neq 0$ for some $\eta^\prime\in X^\ast$.
But $x_{ij}\eta^\prime x_{i_k}\in\supp(e)$ implies that $e$ cannot be palindromic, a contradiction.
So $e$ must have the form of \rref{eq:P-and-E-series}.
\endpr

\begex \label{ex:LTI-palindromic-counter-example}
Suppose $X=\{x_0,x_1\}$.
A  single-input, single-output (SISO) linear time-invariant system is characterized by an impulse response of the form
\begdi
h(t)=\sum_{n=0}^\infty h_n\frac{t^n}{n!}.
\enddi
Repeated application of integration by parts gives
\begin{align*}
y(t)&=\int_0^t h(t-\tau)u(\tau)\,d\tau \\
&=\sum_{n=0}^\infty h_n\int_0^t \frac{(t-\tau)^n}{n!}u(\tau)\,d\tau \\
&=\sum_{n=0}^\infty h_n E_{x_0^nx_1}[u](t) \\
&=F_c[u](t),
\end{align*}
where the generating series is $c=\sum_{n\geq 0} h_n x_0^nx_1$.
In general, a series $c\in\allseries$ is said to be {\em linear} when every word in its support has exactly one instance of the letter $x_j\in X/\{x_0\}$
(recall $u_0:=1$).
In the present context, $\tilde{c}\neq c$. So linear time-invariant systems, in particular, lack all forms of symmetry described above.
On the other hand, linear series whose support contains words of the form $x_0^ix_j x_0^k$ correspond to linear time-varying systems and
can exhibit some forms of symmetry. A particular case is addressed in the next section.
\endex

\section{Globally maximal palindromic SISO linear systems}
\label{sec:Globally Maximal Palindromic Linear Systems}

A palindromic SISO linear system always has an odd palindromic generating series of the form
\begdi 
c=\sum_{n=0}^\infty (c,x_0^nx_1x_0^n)\, x_0^nx_1x_0^n.
\enddi
It is said be {\em locally maximal} when it has the form
\begeq \label{eq:locally-maximal-palindromic-generating-series}
c=\sum_{n=0}^\infty KM^{2n}(2n)!\, x_0^nx_1x_0^n
\endeq
and  {\em globally maximal} when
\begeq \label{eq:globally-maximal-palindromic-generating-series}
c=\sum_{n=0}^\infty KM^{2n}\, x_0^nx_1x_0^n.
\endeq
(For convenience, the constants $K$ and $M$ in Definition~\ref{de:maximal-series} have been redefined here using
the fact that $(c,x_0^nx_1x_0^n)=KM^{2n+1}=(KM)M^{2n}$
and $(c,x_0^nx_1x_0^n)=KM^{2n+1}(2n+1)!=(KM)M^{2n}(2n)!(2n+1)\leq (KM)(2M)^{2n}(2n)!$.)
A Fliess operator $F_c$ is called palindromic when $c$ is palindromic.
In this section, three problems are addressed concerning
globally maximal palindromic SISO linear systems.
First, it is shown that such systems do not have finite Hankel rank or finite Lie
rank. Thus, $F_c$
does not have a finite dimensional bilinear or control affine state space realization.
Next, an explicit form of the system's impulse response is derived from which it is possible to
identify an infinite dimensional state space realization. Finally, the realization's zero dynamics are
described.

\subsection{Nonexistence of finite dimensional realizations}
\label{sec:GC-nonexistent-finite-realizations}

The series \rref{eq:globally-maximal-palindromic-generating-series} is $\re$-algebraic since
it satisfies the single polynomial equation
\begdi
c=Kx_1+M^2x_0cx_0.
\enddi
But the first theorem asserts that $c$ is not rational since its Hankel rank is infinite.
The argument is a minor variation of one given by \citet[p.~283]{Stanley_99} for a
similar type of series.

\begth
The series $c=\sum_{n\geq 0} KM^{2n}x_0^nx_1x_0^n$ has infinite Hankel rank.
\endth

\begpr
If $c$ had finite Hankel rank $n$, then it would have a linear representation $(N_0,N_1,C,z_0)$ of dimension $n$.
From the Cayley--Hamilton theorem, it would follow that for some
set of real numbers $a_i$, $i=1,2,\ldots,n$
\begin{align*}
KM^{2n}&=(c,x_0^nx_1x_0^n)=C N_0^nN_1N_0^nz_0 \\
&=C(a_1N_0^{n-1}+a_2N_0^{n-2}+\cdots+a_nI)N_1N_0^nz_0 \\
&=a_1(c,x_0^{n-1}x_1x_0^n)+a_2(c,x_0^{n-2}x_1x_0^n)+\cdots
+a_n(c,x_1x_0^n)\\
&=0,
\end{align*}
a contradiction. Thus, the Hankel rank of $c$ is not finite.
\endpr

The fact that the Hankel rank of $c$ is infinite does not rule out the possibility that $c$ could have a finite Lie rank, i.e., that
$c$ is a differentially generated series.
But this turns out also not to be the case.

\begth \label{th:GC-finite-Lie-rank}
The series $c=\sum_{n\geq 0} KM^{2n}x_0^nx_1x_0^n$ has infinite Lie rank.
\endth

\begpr
Define the family of polynomials $p_i=\liead_{x_0}^i(x_1)$, $i=0,1,\ldots$ in the free Lie algebra $\liepoly$ using the bracket operation
$[x_0,x_1]:=x_0x_1-x_1x_0$, where $\liead^0_{x_0}(x_1)=x_1$, and
\begdi
\liead_{x_0}^{i+1}(x_1)=[x_0,\liead_{x_0}^{i}(x_1)],\;\;i\geq 0.
\enddi
It is easily verified by induction that
\begdi
p_i=\sum_{k=0}^i {i\choose k} x_0^{i-k}x_1(-x_0)^{k},\;\;i\geq 0,
\enddi
so that for any $\eta\in X^\ast$
\begeq \label{eq:Ad-induction-formula-coefficient}
(p_i,\eta)=\sum_{k=0}^i {i\choose k}(-1)^k (x_0^{i-k}x_1x_0^k,\eta).
\endeq
Given any $\xi\in X^\ast$, it follows from the linearity of ${\mathcal H}_c$ that
\begin{align*}
({\mathcal H}_c(p_i),\xi)&=\sum_{\eta\in X^\ast} ({\mathcal H}_c(\eta),\xi)(p_i,\eta) \\
&=\sum_{\eta\in X^\ast} (c,\xi\eta)(p_i,\eta).
\end{align*}
Setting $\xi=x_0^j$ and substituting \rref{eq:Ad-induction-formula-coefficient} gives
\begin{align*}
({\mathcal H}_c(p_i),x_0^j)&=\sum_{\eta\in X^\ast}\sum_{k=0}^i (c,x_0^j\eta){i\choose k}(-1)^k (x_0^{i-k}x_1x_0^k,\eta)\\
&=\sum_{k=0}^i (c,x_0^{i+j-k}x_1x_0^k){i\choose k}(-1)^k\\
&=\left\{
\begin{array}{ccl}
KM^{2k}{i\choose k}(-1)^k & : & 2k=i+j,\;\; j\leq i \\
0 & : & \mbox{otherwise}.
\end{array}
\right.
\end{align*}
Thus, every ${\mathcal H}_c(p_i)$ is a polynomial in $\allpolyXO$, where $X_0=\{x_0\}$.
For example,
\begin{align*}
{\mathcal H}_c(p_0)&=K \\
{\mathcal H}_c(p_1)&=-KM^2x_0 \\
{\mathcal H}_c(p_2)&=-2KM^2+KM^4x_0^2 \\
{\mathcal H}_c(p_3)&=3KM^4x_0-KM^6x_0^3 \\
&\hspace*{0.09in}\vdots
\end{align*}
In particular, for any $i\geq 0$,
$
({\mathcal H}_c(p_i),x_0^i)=KM^{2i}(-1)^i\neq 0.
$
Therefore, ${\mathcal H}_c(p_i)$ is a polynomial of degree $i$.
Let $S_p$ denote the subspace of $\liepoly$ spanned by the $p_i$'s. Necessarily,
$\dim({\mathcal H}_c(S_p))\leq \dim({\mathcal H}_c(\liepoly))$.
But ${\mathcal H}_c(S_p)$ is not a finite dimensional subspace of $\allpolyXO$.
Thus, ${\mathcal H}_c(\liepoly)$ cannot be finite dimensional.
\endpr

\subsection{An infinite dimensional realization}
\label{sec:Infinite-Dimensional-State-Space-Realization}

\begin{figure}[tb]
\begin{center}
\includegraphics[scale=0.3]{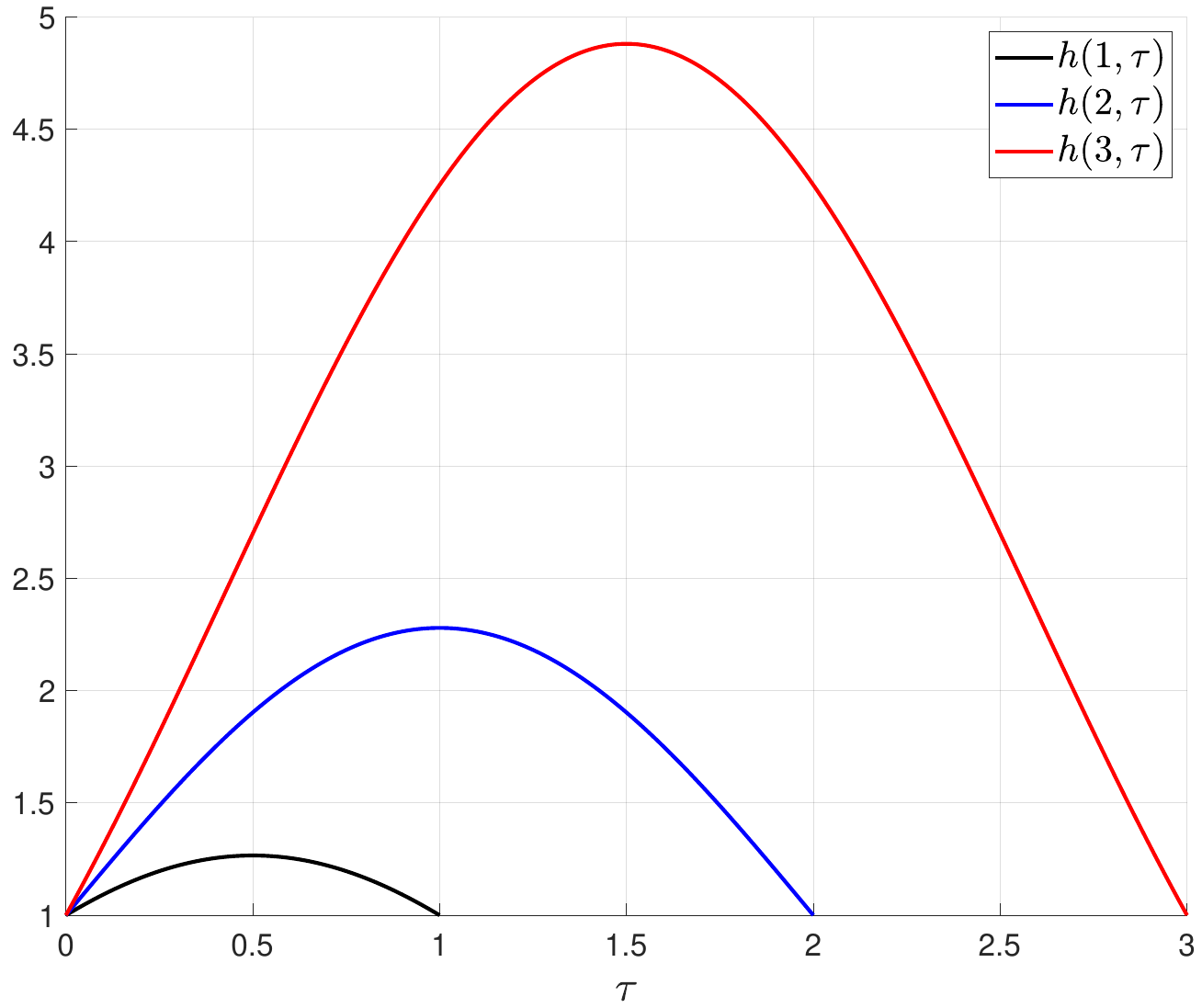}
\end{center}
\caption{Plot of $h(t,\tau)$ versus $\tau$ for $t=1,2,3$ and $K=M=1$ in the globally maximal case}
\label{fig:Bessel-plot}
\end{figure}

The palindromic system $y=F_c[u]$, where $c$ is given by \rref{eq:globally-maximal-palindromic-generating-series}
can be written in the form
\begdi 
y(t)=\int_0^t h(t,\tau)u(\tau)\,d\tau,
\enddi
where the impulse response is
\begeq \label{eq:h-palindrome-series}
h(t,\tau)=K\sum_{n=0}^\infty M^{2n}\frac{(t-\tau)^n}{n!}\frac{\tau^{n}}{n!}
\endeq
for $t\geq\tau\geq 0$.
Observe that \rref{eq:h-palindrome-series} can be written in the form
\begeq \label{eq:h-Bessel-series}
h(t,\tau)=K\sum_{n=0}^\infty\frac{(Mt)^n}{n!}J_n(2M\tau),
\endeq
where
\begeq \label{eq:Jn-series}
J_n(\tau)=\sum_{k=0}^\infty \frac{(-1)^k}{k!(n+k)!}\left(\frac{\tau}{2}\right)^{n+2k}
\endeq
is the $n$-th order Bessel function of the first kind.
Thus, the input-output map is
\begeq \label{eq:palindrome-IO-Jn}
y(t)=K\sum_{n=0}^\infty \frac{(Mt)^n}{n!}\int_0^t J_n(2M\tau) u(\tau)\,d\tau.
\endeq
Using the
Bessel function multiplication identity,
\begdi
J_0(\lambda z)=\sum_{n=0}^\infty \frac{1}{n!} \left(\frac{(1-\lambda^2)z}{2}\right)^n J_n(z),\;\;\forall z,\lambda\in\mathds C
\enddi
(e.g., see \cite{Truesdell_50}),
equation \rref{eq:h-Bessel-series} simplifies to
\begdi
h(t,\tau)=KJ_0(2iM\sqrt{(t-\tau)\tau})=KI_0(2M\sqrt{(t-\tau)\tau}), 
\enddi
where $I_0(z)$ denotes the zeroth order modified Bessel function of the first kind (see Figure~\ref{fig:Bessel-plot}).
From \rref{eq:palindrome-IO-Jn} it follows directly that $F_c$
has the infinite dimensional, linear time-varying realization
\begin{subequations}\label{eq:LTV-palindromic-system}
\begin{align}
\dot{z}(t)&=B(t)u(t),\;\;z(0)=0 \\
y(t)&=C(t)z(t)
\end{align}
\end{subequations}
with $z=[z_0 \; z_1 \; z_2 \cdots]^T$,
$B=[B_0 \; B_1\;B_2 \cdots]^T$, and C=$[C_0 \; C_1 \; C_2 \cdots]$,
where $z_n(t)=\int_0^t J_n(2M\tau) u(\tau)\,d\tau$, $B_n(t)=J_n(2Mt)$, and $C_n(t)=K(Mt)^n/n!$.

Next, observe that
\begdi
\dot{y}(t)=KM\sum_{n=0}^\infty \frac{(Mt)^n}{n!}z_{n+1}(t)
+\left(K\sum_{n=0}^\infty \frac{(Mt)^n}{n!} J_n(2Mt)\right) u(t).
\enddi
Since $J_0(0)=1$, it follows that system \rref{eq:LTV-palindromic-system} has relative degree $r=1$ at every point
$z_0\in\re^{\infty}$. Thus, if it is initialized at any $z(0)$ whose first component $z_{0}(0)=0$,
then this system has zero dynamics of the form
\begeq \label{eq:GC-palindromic-zero-dynamics}
\dot{z}_n^\ast=J_n(2Mt)u^\ast(t),\;\;z^\ast_n(0)=\frac{z_{0,n}}{n!},\;\;n\geq 1,
\endeq
where
\begdi
u^\ast(t)=-\frac{M\sum_{n=0}^\infty \frac{(Mt)^n}{n!}z^\ast_{n+1}(t)}{\sum_{n=0}^\infty \frac{(Mt)^n}{n!} J_n(2Mt)}.
\enddi
A more explicit expression for these dynamics is developed in
the next section.

\subsection{Zero dynamics}
\label{sec:Zeroing-the-output}

Consider first the general problem of selecting an input $u^\ast$ such that the output
$y=F_c[u^\ast]$ is exactly the zero function on an interval $[0,T]$ over which the series converges.
In the SISO case where $X=\{x_0,x_1\}$, the generating series $c\in\allseries$ can be partitioned into its natural part $c_N=\sum_{k\geq 0} (c,x_0^k)x_0^k$ and its forced part $c_F=c-c_N$. Note that every word in the support of $c_N$ is a palindrome.
Let $r\geq 1$ be the largest integer such that $\supp(c_F)\subset x_0^{r-1}X^\ast$. Then $c$ is said to have {\em relative degree} $r$ if the word $x_0^{r-1}x_1$ is also in
the support of $c_F$. Otherwise, $c$ does not have relative degree \citep{Gray-etal_14}. This notion is consistent with the state space definition of relative degree when $F_c$ has a control-affine realization.
The series $c$ is said to be {\em primely nullable} when there exists a unique input $u^\ast$ such that the output $y=F_c[u^\ast]$ is zero on some
interval $[0,T]$. A sufficient condition for $c$ to be primely nullable is having relative degree $r$ and $\supp(c_N)\subseteq x_0^rX_0^\ast$ being nonempty \citep[Theorem 3.1]{Gray-etal_24}.

Consider now the case where $c_F=\sum_{n\geq 0} KM^{2n}x_0^nx_1x_0^n$, which has relative degree one as expected.
The series $c=c_N+c_F$ is then primely nullable for any $c_N=\sum_{k\geq 1} (c,x_0^k)x_0^k$.
The following theorem is essential.

\begth \label{th:zero-output-GC-palindrome}
Let $c_N=\sum_{k\geq 1} (c,x_0^k)x_0^k$ be a globally convergent series.
The palindromic series $c=c_N+\sum_{n\geq 0} KM^{2n} x_0^nx_1x_0^n$ satisfies $F_c[u^\ast]=0$ on any finite interval $[0,T]$, where $tu^\ast(t)$ has the
Neumann series in terms of Bessel functions
\begdi
tu^\ast(t)=\sum_{k=1}^\infty -(c,x_0^k)\frac{k}{KM^k}J_k(2Mt),
\enddi
or equivalently,
\begeq \label{eq:ustar-eqn-GC-maximal-case}
u^\ast(t)=\sum_{k=1}^\infty (c,x_0^k)u_k^\ast(t),
\endeq
where $u_k^\ast(t):= -kJ_k(2Mt)/(KM^kt)$.
\endth

\begin{figure}[tb]
\begin{center}
\includegraphics[scale=0.3]{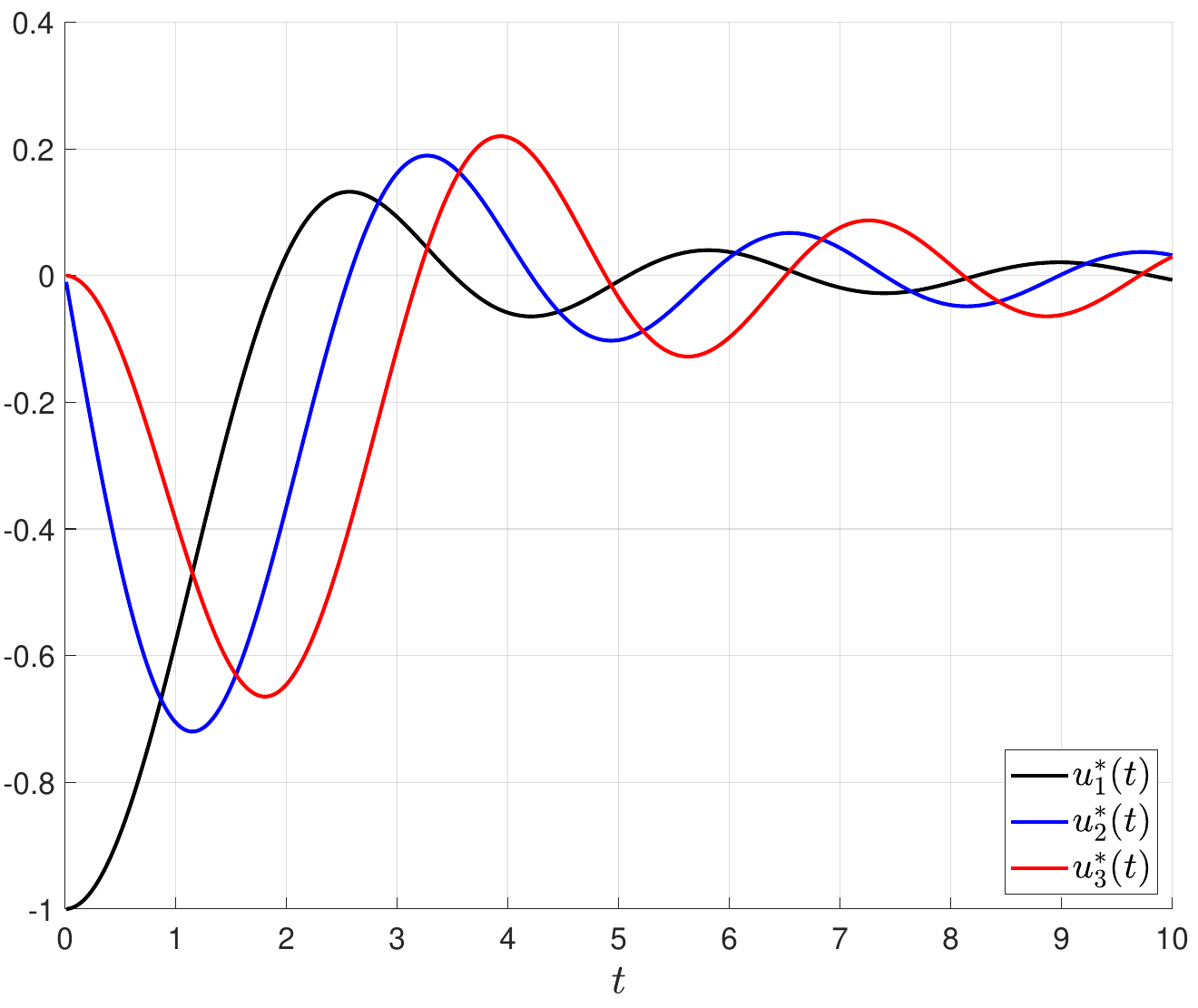}
\end{center}
\caption{Input $u_k^\ast(t)=-kJ_k(2t)/t$, $k=1,2,3$}
\label{fig:ustar-plot}
\end{figure}

\begin{figure}[tb]
\begin{center}
\includegraphics[scale=0.35]{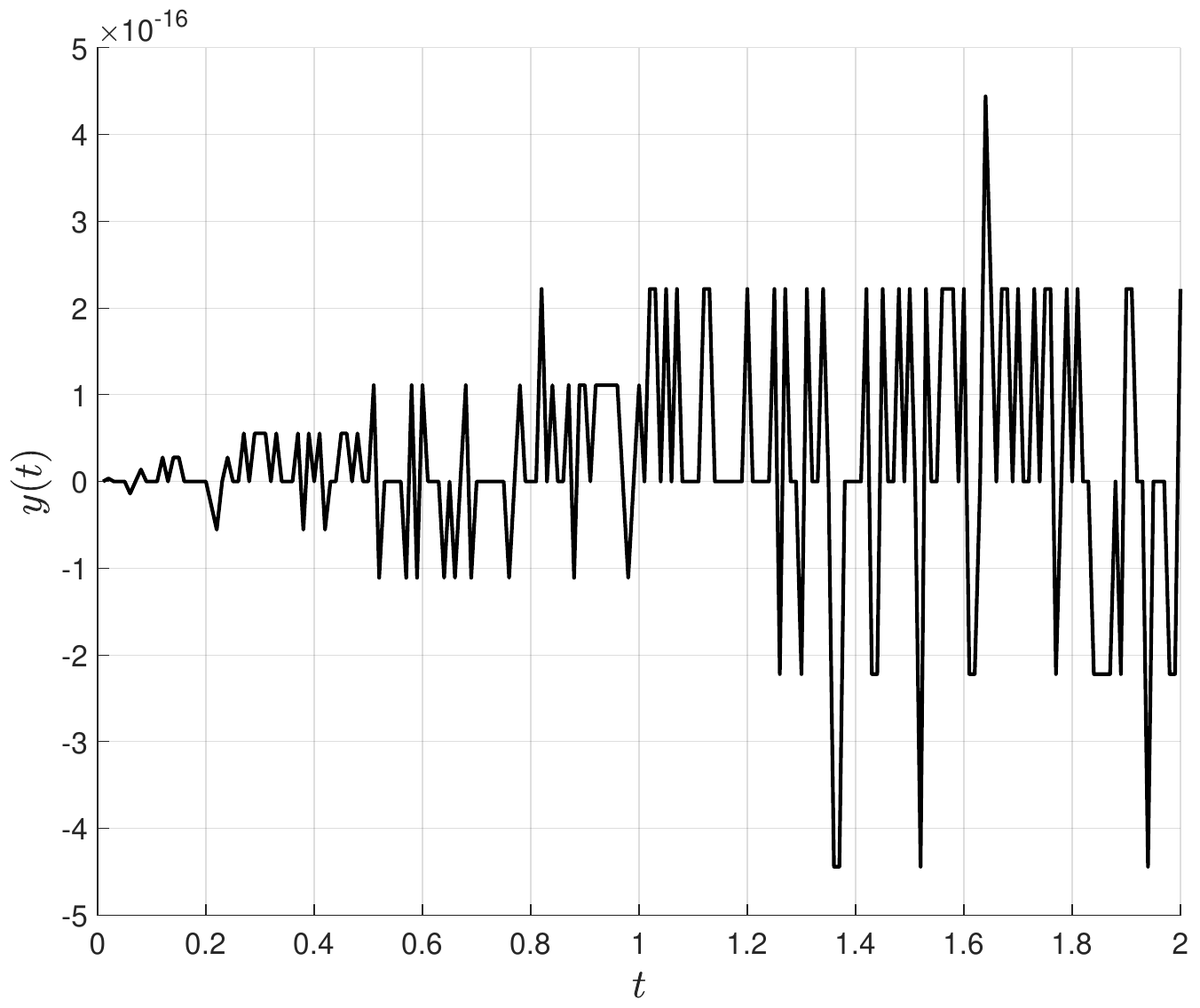}
\end{center}
\caption{Simulated output $y=F_{x_0+c_F}[u_1^\ast]$ when $K=M=1$}
\label{fig:zero-output-plot}
\end{figure}

The first three $u_k^\ast$ are shown in Figure~\ref{fig:ustar-plot} when $K=M=1$. For the simplest case where $c_N=x_0$, the numerically computed output $y(t)$ is equivalent to machine zero for all $t\geq 0$ as shown
in Figure~\ref{fig:zero-output-plot}. In fact, this is the case for {\em every} simulated output $F_{x_0^k+c_F}[u_k^\ast]$, $k\geq 1$. This
motivates the following lemma from which the proof of the theorem follows directly from superposition.

\begle \label{le:GC-palindromic-ustar}
Fix real number $M>0$. For any integer $k\geq 1$,
\begdi
\frac{(Mt)^k}{k!}-\int_0^t  I_0(2M\sqrt{(t-\tau)\tau})\frac{k}{\tau}J_k(2M\tau)\,d\tau=0,\;\;t\geq 0.
\enddi
\endle

\begpr
The series expansions \rref{eq:Jn-series} and
\begeq \label{eq:In-series}
I_0(z)=\sum_{i=0}^\infty \frac{1}{(i!)^2}\left(\frac{z}{2}\right)^{2i}
\endeq
are first applied to yield a series expansion for the integral
\begin{align*}
\lefteqn{\int_0^t I_0(2M\sqrt{(t-\tau)\tau})\frac{1}{M\tau}J_k(2M\tau)\,d\tau} \hspace*{0.4in} \\
&=\sum_{i,j=0}^\infty \frac{(-1)^j M^{2i+2j+k}}{(i!)^2j!(j+k)!} \int_0^t(t-\tau)^i\tau^{i+2j+k-1}\,d\tau.
\end{align*}
The change of variables $\tau=t \tau^\prime$ yields the expression
\begin{align*}
\lefteqn{\int_0^t I_0(2\sqrt{(t-\tau)\tau})\frac{1}{\tau}J_k(2\tau)\,d\tau}\hspace*{0.18in} \\
&=\left[\sum_{i,j=0}^\infty \frac{(-1)^j (Mt)^{2i+2j}}{(i!)^2j!(j+k)!} B(i+2j+k,i+1)\right]\hspace*{-0.05in} (Mt)^k,
\end{align*}
where
\begdi
B(z_1,z_2)=\int_0^1 \tau^{z_1-1}(1-\tau)^{z_2-1}\,d\tau
\enddi
is the beta function. From the identity
\begdi
B(m,n)=\frac{(m-1)!(n-1)!}{(m+n-1)!},\;\;\forall m,n\in\nat,
\enddi
it follows that the term in the square brackets above is
\begdi
\sum_{i,j=0}^\infty (-1)^j \frac{(i+2j+k-1)!}{i!j!(j+k)!(2i+2j+k)!} (Mt)^{2i+2j}.
\enddi
The assertion is that this infinite sum is equal to $1/(k(k!))$ for all $k\geq 1$ and does not depend on the value of $M$ or $t$.
The proof is technical. It can be found in Appendix A. Using this identity,
the theorem is proved.
\endpr

Returning to the problem of determining the zero dynamics, combining \rref{eq:GC-palindromic-zero-dynamics} and \rref{eq:ustar-eqn-GC-maximal-case}
yields for $n\geq 1$
\begdi
\dot{z}_n^\ast(t)=-J_n(2Mt)\sum_{k=1}^\infty (c,x_0^k)\frac{k J_k(2Mt)}{KM^k t} ,\;\;t\geq 0
\enddi
with $z_n(0)=(c,x_0^n)/n!$.

\begex \label{ex:zero-output-globally-maximal-palindromic-system}
Suppose $c_N=\sum_{k\geq 1} (-1)^{k+1} 2k\, x_0^{2k}$ and $c_F=\sum_{n\geq 0} x_0^nx_1x_0^n$. Then
\begdi
F_{c_N}[u](t)=\sum_{k=1}^\infty (-1)^{k+1} 2k\, \frac{t^{2k}}{(2k)!}=2t\sin(t),\;\;t\geq 0
\enddi
and
\begin{align*}
u^\ast(t)&=\sum_{k=1}^\infty (-1)^{k+1} 2k \left( -\frac{2k}{t}J_{2k}(2t)\right) \\
&= -\sin(2t),\;\;t\geq 0
\end{align*}
using the known identity
\begdi
\sum_{k=1}^\infty (-1)^{k+1}(2k)^2 J_{2k}(z)=\frac{z\sin(z)}{2}
\enddi
\citep[p. 974]{Gradshteyn-Ryzhik_80}.
Again, a numerical simulation shows that $y(t)$ is equivalent to machine zero for all $t\geq 0$.
The zero dynamics must satisfy
\begdi
\dot{z}_n^\ast(t)=J_n(2t)\sin(2t) ,\;\;t\geq 0.
\enddi
with $z_n(0)=0$ for $n\geq 1$ odd and $z_n(0)=-(-1)^{n/2}/(n-1)!$ for $n\geq 2$ even.
It is not difficult to show that none of these states are bounded as $t$ goes to infinity.
So the system's zero dynamics are not stable in any sense.
\endex

\section{Locally maximal palindromic SISO linear systems}
\label{sec:Locally Maximal Palindromic Linear Systems}

In this section, the same three problem from the previous section are
addressed except now for locally maximal palindromic SISO linear systems.
The main difference is that here the impulse response has a singularity,
and the radius of convergence of the Fliess operator is finite.

\subsection{Nonexistence of finite dimensional realizations}

The first theorem rules out the possibility that locally maximal palindromic linear systems are
rational, i.e., have finite Hankel rank.

\begth
The series $c=\sum_{n\geq 0} KM^{2n}(2n)!\,x_0^nx_1x_0^n$ has infinite Hankel rank.
\endth

\begpr
The proof is by contradiction. If $c$ had finite Hankel rank, then it would have a linear representation
$(C,N_0,N_1,z_0)$ so that \rref{eq:linear-representation} is satisfied. This would imply that
\begdi
\abs{(c,x_{i_k}\cdots x_{i_1})}=|CN_{i_k}\cdots N_{i_i}z_0|.
\enddi
Applying the Cauchy-Schwarz inequality and using any submultiplicative matrix norm, it would then follow that
\begdi
\abs{(c,x_{i_k}\cdots x_{i_1})}\leq\norm{C}\norm{N_{i_k}}\cdots \norm{N_{i_i}}\norm{z_0}\leq KM^k,
\enddi
where $K=\norm{C}\norm{z_0}$ and $M=\max(\norm{N_0},\norm{N_1})$. This is clearly a contradiction as the coefficients of $c$ are
growing at a factorial rate.
\endpr

It is conjectured that $c$ above is also not algebraic for the same reason that the coefficients are growing
faster than what a proper $\re$-algebraic system can produce. A precise argument would require using the representation theory
of Shamir for algebraic series \citep{Petre-Salomaa_09,Shamir_67}, but that issue is beyond the scope of the present work.
The next theorem is the local version of  Theorem~\ref{th:GC-finite-Lie-rank}, which again eliminates the possibility
of a control-affine realization.

\begth
The series $c=\sum_{n\geq 0} KM^{2n}(2n)!\,x_0^nx_1x_0^n$ has infinite Lie rank.
\endth

\begpr
The same proof for Theorem~\ref{th:GC-finite-Lie-rank} applies in this case.
\endpr

\subsection{An infinite dimensional realization}

\begin{figure}[tb]
\begin{center}
\includegraphics[scale=0.3]{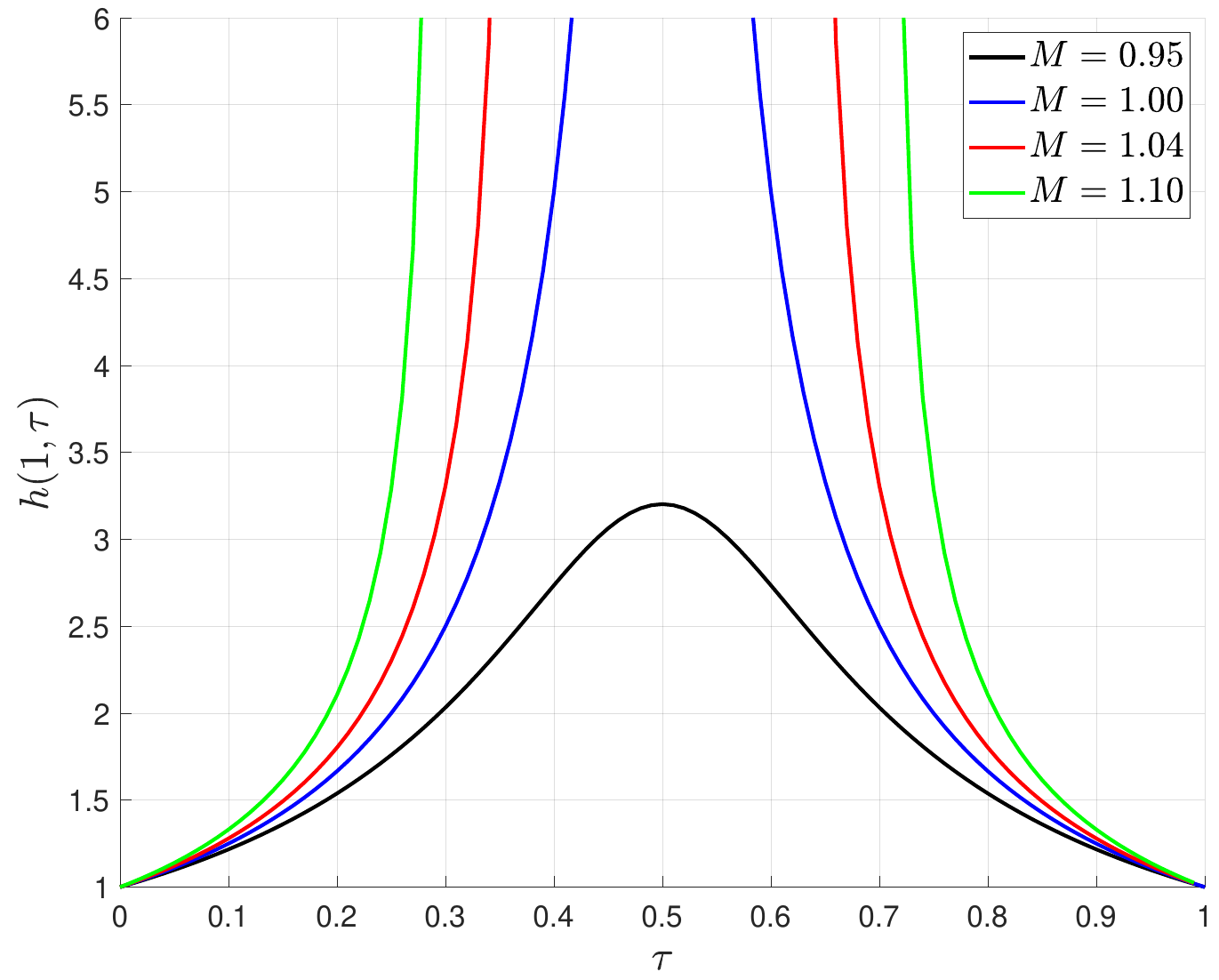}
\end{center}
\caption{Plot of $h(1,\tau)$ versus $\tau$ for $K=1$ and various $M$ in
the locally maximal case}
\label{fig:elliptic-plot}
\end{figure}

The impulse response for \rref{eq:locally-maximal-palindromic-generating-series} is computed
directly from \rref{eq:h-palindrome-series} by introducing the factorial term $(2n)!$ from the
coefficients, that is, for $t\geq \tau\geq 0$
\begin{align}
h(t,\tau)&=K\sum_{n=0}^\infty \frac{(2n)!}{n!}(M^2t)^n J_n(2M\tau) \nonumber\\
&=\frac{K}{\sqrt{1-4M^2(t-\tau)\tau}}, \label{eq:h-LC-case}
\end{align}
which has two distinct real singularities whenever $Mt>1$, namely at
$\tau_\pm=(t\pm p(t))/2$ with $p(t):=\sqrt{(Mt)^2-1}/M$ (see Figure~\ref{fig:elliptic-plot}).
For the case where $Mt<1$, the presence of complex singularities $\tau_\pm$ will bound the radius of convergence
of the Taylor series of $h(t,\cdot)$ about $\tau=0$ by $\abs{\tau_{\pm}}$.
The corresponding state space model is the same as that given in \rref{eq:LTV-palindromic-system}
except now $C_n(t)=K(MT)^n(2n)!/n!$, $n\in\nat_0$. Therefore, this realization also has
relative degree $r=1$ and its zero dynamics are still given by \rref{eq:GC-palindromic-zero-dynamics}
except that $u^\ast$ has to be recomputed as presented in the next section.

\subsection{Zero dynamics}

Given two vectors $a\in\re^p$ and $b\in\re^q$, the {\em generalized hypergeometric function}
\begdi
_p\hspace*{-0.01in}F_{q}(a;b;z)=\sum_{k=0}^\infty \frac{(a_1)_k\cdots (a_p)_k}{(b_1)_k\cdots(b_q)_k}\frac{z^k}{k!}
\enddi
is defined on the complex plane $\C$ with
$(x)_k:=\Gamma(x+k)/\Gamma(x)$ being the Pochhammer symbol, and $\Gamma$ is the gamma function  \citep{Askey-Olde_Daalhuis_10}.
In general, $_{q+1}\hspace*{-0.01in}F_{q}(a;b;z)$ is a multi-valued function
with a branch cut discontinuity in the complex plane running from $-\infty$ to -1 and from 1 to $\infty$.
It will be assumed here that such functions are defined in terms of their principal values when $z\in [0,1)$.
In light of \rref{eq:Jn-series} and \rref{eq:In-series}, it is not difficult to show that for any nonnegative real
number $\alpha$:
\begin{align*}
J_{\alpha }(x)&={}_{0}F_{1}\left(;\alpha +1;-{\tfrac {1}{4}}x^{2}\right){\frac {({\tfrac {x}{2}})^{\alpha }}{\Gamma (\alpha +1)}} \\
I_{\alpha }(x)&={}_{0}F_{1}\left(;\alpha +1;{\tfrac {1}{4}}x^{2}\right){\frac {({\tfrac {x}{2}})^{\alpha }}{\Gamma (\alpha +1)}}.
\end{align*}
The function $_{0}F_{1}(;a;z)$ is a limiting case usually referred to as a  {\em confluent hypergeometric function}.
Given these identities and Theorem~\ref{th:zero-output-GC-palindrome}, the following theorem is not entirely unexpected.

\begth \label{th:zero-output-LC-palindrome}
Let $c_N=\sum_{k\geq 1} (c,x_0^k)x_0^k$ be a locally convergent series so that $y_N=F_{c_N}[u]$ converges on $[0,T_N]$.
Given the palindromic series $c_F=\sum_{n\geq 0} KM^{2n} (2n)!\,x_0^nx_1x_0^n$, define
\begin{align*}
f^e_{j}(t)&=K \,_3\hspace*{-0.005in}F_{2}(a^e_j;b^e_j;(Mt)^2)\frac{t^{1+2j}}{1+2j},\;\;j\geq 0 \\
f^o_0(t)&=K\, _2\hspace*{-0.005in}F_{1}(a^o_0;b^o_0;(Mt)^2) \frac{t^2}{2} \\
f^o_{j}(t)&=K\, _3\hspace*{-0.005in}F_{2}(a^o_j;b^o_j;(Mt)^2)\frac{t^{2+2j}}{2+2j},\;\;j\geq 1,
\end{align*}
where
\begin{align*}
a^e_j&=\left[\frac{1}{2},\;1,\;1+2j\right],\;\; b^e_j=\left[1+j,\;\frac{3}{2}+j\right],\;\;j\geq 0 \\
a^o_0&=\left[1,\frac{1}{2}\right],\;\;b^o_0=\left[\frac{3}{2}\right] \\
a^o_j&=\left[\frac{1}{2},\;1,\;2+2j\right],\;\; b^o_j=\left[\frac{3}{2}+j,\;2+j\right],\;\;j\geq 1.
\end{align*}
If $MT<1$, and $u^\ast$ is a power series whose coefficients $u^\ast_j$, $j\geq 0$ satisfy
\begeq \label{eq:ustar-eqn-LC-maximal-case}
y_N(t)+\sum_{j=0}^\infty u^\ast_{2j} f^e_j(t)+u^\ast_{2j+1}f^o_j(t)=0,
\endeq
then $y=F_c[u^\ast]=0$ on $[0,\min(T_N,T)]$.
\endth

\begpr
A straightforward calculation gives for $j\geq 0$
\begin{align*}
\int_0^t h(t,\tau)\tau^{2j}\,d\tau
&=\sum_{n=j}^\infty \frac{(2(n- j))!\,(n+j)!}{(n-j)!}M^{2(n-j)} \frac{t^{1+2n}}{(1+2n)!} \\
&=K \,_3\hspace*{-0.005in}F_{2}(a^e_j;b^e_j;(Mt)^2)\frac{t^{1+2j}}{1+2j},
\end{align*}
where $h$ is given by \rref{eq:h-LC-case} and $t\in[0,T]$ with $MT<1$. Likewise,
\begin{align*}
\int_0^t h(t,\tau)\tau\,d\tau
&= \sum_{n=0}^\infty \frac{(2n)!}{(2n+1)!}M^{2n}\frac{t^{2n+2}}{2}\\
&=K\, _2\hspace*{-0.003in}F_{1}(a^o_0;b^o_0;(Mt)^2)\frac{t^2}{2}
\end{align*}
and
\begin{align*}
\int_0^t h(t,\tau)\tau^{1+2j}\,d\tau
&=\sum_{n=j}^\infty \frac{(2(n-j))!\,(n+j+1)!}{(n-j)!\,(n+1)} M^{2(n-j)} \frac{t^{2+2n}}{2(1+2n)!} \\
&=K\, _3\hspace*{-0.005in}F_{2}(a^o_j;b^o_j;(Mt)^2)\frac{t^{2+2j}}{2+2j}.
\end{align*}
Setting $u^\ast(\tau)=\sum_{j=0}^\infty u^\ast_j \tau^j$, it then follows
from linearity that the forced response is
\begdi
y_F(t)=\sum_{j=0}^\infty u^\ast_{2j} f^e_j(t)+u^\ast_{2j+1}f^o_j(t).
\enddi
If the coefficients are then selected to satisfy \rref{eq:ustar-eqn-LC-maximal-case}, the system
output is $y=y_N+y_F=0$ on $[0,\min(T_N,T)]$ as desired.
\endpr

Equation~\rref{eq:ustar-eqn-LC-maximal-case} can be written in the form of a
linear set of equations $Ax=b$ by equating the coefficients of like powers of $t$.
In this case,
$A$ turns out to be lower triangular with no zeros along the diagonal. In contrast,
Theorem~\ref{th:zero-output-GC-palindrome} for the global case corresponds to a system where $A$ is
diagonal, hence the more explicit expression \rref{eq:ustar-eqn-GC-maximal-case} for $u^\ast$ is possible.
In the present context then the zero dynamics are given by
\begdi
\dot{z}_n^\ast(t)=J_n(2Mt)\sum_{k=0}^\infty u_k^\ast ,\;\;t\geq 0
\enddi
with $z_n(0)=(c,x_0^n)/n!$, $n\geq 1$, and where the coefficients of $u^\ast$ satisfy \rref{eq:ustar-eqn-LC-maximal-case}.
Of course, the input series yielding the zero output of the system only converges on a finite interval as described in
Theorem~\ref{th:zero-output-LC-palindrome}.

\begin{figure}[tb]
\begin{center}
\includegraphics[scale=0.3]{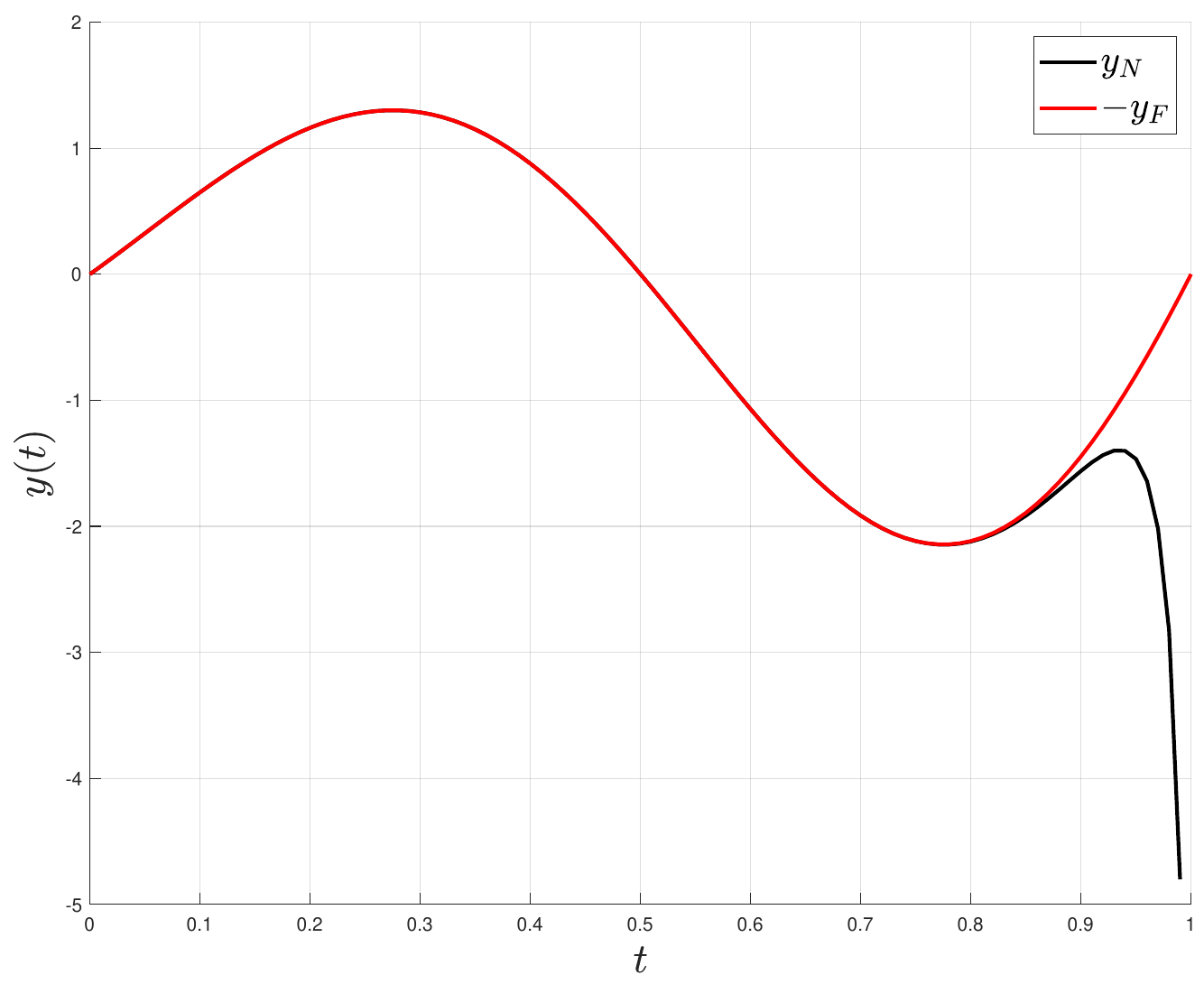}
\end{center}
\caption{Plot of $y_N$ and $-y_F$ in Example~\ref{ex:zero-output-LC-case}}
\label{fig:yN-yF-plot-LC}
\end{figure}

\begex
Consider a locally maximal palindromic linear system $c=c_F+c_N$, where
\begin{align*}
c_F&=\sum_{n\geq 0}KM^{2n}(2n)!\,x_0^nx_1x_0^n \\
c_N&=\sum_{k=1}^\infty M^{2k-1} (2(k-1))!\, x_0^{2k-1}.
\end{align*}
Here $y_N$ is the odd function ${\rm Arctanh}(Mt)$ converging on $[0,1)$.
This example is simple enough that \rref{eq:ustar-eqn-LC-maximal-case}
can be solved directly to give $u^\ast(t)=-M/K$, $t\in[0,1)$. In which case,
the zero dynamics are given by
\begdi
z^\ast_n(t)=-\frac{M}{K}J_n(2Mt),\;\;t\in [0,1)
\enddi
so that
\begin{align*}
z_n(t)
&=z_n(0)-\frac{M}{K}\int_0^t J_n(2M\tau)\,d\tau \\
&=\frac{M^n}{n}\delta_n^o-\frac{1}{K}\,
{}_1\hspace*{-0.005in}F_{2}\left(\left[\frac{1}{2}+\frac{n}{2}\right];\left[\frac{3}{2}+\frac{n}{2},1+n\right];-(Mt)^2\right)
\frac{(Mt)^{1+n}}{(1+n)!},
\end{align*}
where $\delta_n^o=1$ when $n$ is odd and zero otherwise. Given the finite interval of
convergence, a stability analysis of the zero dynamics is not possible.
\endex

\begex
\label{ex:zero-output-LC-case}
Consider next the case of a
locally maximal palindromic linear system with
$K=1$, $M=1$, and
$y_N(t)={\rm e}^t\sin(2\pi t)$.
This function has a globally convergent generating series $c_N$ which is neither even or odd, so the
problem of computing $u^\ast$ is more complex.
It can be solved approximately by truncating \rref{eq:ustar-eqn-LC-maximal-case}
to a degree 20 polynomial and then solving numerically. The Mathematica code to do this
is given in Appendix B. Figure~\ref{fig:yN-yF-plot-LC} shows the output components $y_N$ and $-y_F$,
where the latter is computed numerically using \rref{eq:h-LC-case} in the convolution integral.
Ideally, these components should match exactly
so that $y=y_N+y_F=0$. But the truncation error in $y_F$ becomes apparent as $t$ approaches the convergence boundary at $T=1$.
Here the zero dynamics can be determined numerically, but they do not have a simple expression as in the previous example.
\endex

\section{Conclusions}

Two new notions of symmetry for Chen--Fliess input-output systems were given:
coefficient reversible symmetry and palindromic symmetry. Each concept was then related to an existing type of symmetry
described by exchangeable generating series.
Next, a detailed analysis was given for
globally and locally maximal palindromic SISO linear systems.
In each case, it was shown that such generating series
have an infinite
Hankel rank and Lie rank, have a certain infinite dimensional state space realization, and a description of their relative degree and zero dynamics was given.

\section*{Acknowledgment}
The second author thanks the Mathematisches Institut f\"{u}r Mathematik, Julius Maximilians
Universit\"{a}t W\"{u}rzburg, W\"{u}rzburg, Germany
for its support and hospitality while part of this work was performed.

\appendix

\section{Combinatorial identity}
\label{app:Combinatorial-Identity}

\begle
For all $M,t\in \re$ and integers $k\geq 1$
\begdi
\sum_{i,j=0}^\infty (-1)^j \frac{(i+2j+k-1)!}{i!j!(j+k)!(2i+2j+k)!} (Mt)^{2i+2j}=\frac{1}{k(k!)}.
\enddi
\endle

\begpr
Setting $\ell=i+j$ on the left-hand side gives
\begdi
\sum_{\ell=0}^\infty \left[\sum_{j=0}^\ell (-1)^j \frac{(\ell+j+k-1)!}{(\ell-j)!j!(j+k)!}\right] \frac{(Mt)^{2\ell}}{(2\ell+k)!}.
\enddi
The first term in the infinite series where $\ell=0$ is $1/(k(k!))$.
Thus, the claim is that
\begdi
\sum_{j=0}^\ell (-1)^j \frac{(\ell+j+k-1)!}{(\ell-j)!j!(j+k)!}=0,\;\;\forall \ell>0,
\enddi
or equivalently,
\begdi
\sum_{j=0}^\ell (-1)^j {\ell+k+j-1 \choose j}{\ell+k \choose \ell-j}=0, \;\;\forall \ell>0.
\enddi
From upper negation \cite[p.~164]{Graham-etal_94}, it follows that
\begdi
(-1)^j {\ell+k+j-1 \choose j}={-(\ell+k)\choose j}.
\enddi
Therefore, using the Vandermonde convolution identity \cite[p.~174]{Graham-etal_94}
\begin{align*}
\lefteqn{\sum_{j=0}^\ell (-1)^j {\ell+k+j-1 \choose j}{\ell+k \choose \ell-j}} \hspace*{0.4in} \\
&= \sum_{j=0}^\ell {-(\ell+k)\choose j}{\ell+k \choose \ell-j} \\
&={0 \choose \ell}=0
\end{align*}
as claimed when $\ell>0$.
\endpr

\section{Code to compute $u^\ast$: locally maximal case}

The following Mathematica (v12) code was used to
compute an estimate of
$u^\ast$ using \rref{eq:ustar-eqn-LC-maximal-case} with the upper bound truncated to
$J=10$ and where $y_N(t)={\rm e}^t\sin(2\pi t)$.

{\tt
\begin{verbatim}
K=1; M=1;J=10;
aej={1/2,1,1+2j}; bej={1+j,3/2+j};
fe[j_]=K HypergeometricPFQ[aej,bej,(M t)^2]
   t^(1+2j)/(1+2j);
ao0={1,1/2};bo0={3/2};
fo0=K HypergeometricPFQ[ao0,bo0,(M t)^2] t^2/2;
aoj={1/2,1,2+2j}; boj={3/2+j,2+j};
fo[j_]=K HypergeometricPFQ[aoj,boj,(M t)^2]
   t^(2+2j)/(2+2j);
feSum=Sum[ustarcoef[2j]*fe[j],{j,0,J}];
foSum=Sum[ustarcoef[2j+1]*fo[j],{j,1,J}];
yN=Exp[t]Sin[2 Pi t];
sum=Series[yN+feSum+ustarcoef[1]fo0
   +foSum,{t,0,1+2J}];
sol=Solve[CoefficientList[sum,t]==0,
   Array[ustarcoef,1+2J,0]];
ustar[t_] = Sum[ustarcoef[j]*t^j, {j, 0, 2 J}]
   /. sol
\end{verbatim}
}

\end{document}